\documentclass[twocolumn,showpacs,amsfonts,aps,prc,nofootinbib,floatfix]{revtex4}

\usepackage{bm}
\usepackage{graphicx}
\usepackage{amsmath}

\def\Np{N_\mathrm{part}}
\def\ve{\varepsilon}

\def\eq{{\,=\,}}

\usepackage{ulem}
\usepackage{color}

\begin{document}


\title{Hydrodynamic event-plane correlations in Pb+Pb collisions at
$\bm{\sqrt{s}\eq2.76\,A}$\,TeV}

\author{Zhi Qiu}
\author{Ulrich Heinz}
\affiliation{Department of Physics, The Ohio State University,
  Columbus, Ohio 43210-1117, USA}

\begin{abstract}
The recently measured correlations between the flow angles associated with higher harmonics in the anisotropic flow generated in relativistic heavy-ion collisions are shown to be of hydrodynamic origin. The correlation strength is found to be sensitive to both the initial conditions and the shear viscosity of the expanding fireball medium.
\end{abstract}


\date{\today}

\maketitle

{\sl 1.\;Introduction.}
Due to the fluctuating positions of the nucleons inside the colliding nuclei at the point of impact
\cite{Miller:2003kd} and to quantum fluctuations of the quark and gluon fields inside those nucleons \cite{Dusling:2011rz,Muller:2011bb,Dumitru:2012yr,Schenke:2012wb}, the initial 
density profiles of the fireballs created in relativistic heavy-ion collisions fluctuate in size, shape and magnitude from event to event, even for collisions with identical impact parameters. One way to characterize these fluctuating initial profiles is through a set of harmonic eccentricity coefficients $\ve_n$ with associated ``participant plane" angles $\Phi_n$ (see, e.g., 
\cite{Alver:2010gr,Qin:2010pf,Teaney:2010vd}):
\begin{eqnarray} 
\label{eq:1}
  &&\ve_1\,e^{i \Phi_1}
  = -\frac{\int r\,dr\,d\phi\, r^3 e^{i \phi}\,e(r,\phi)}{\int r\,dr\,d\phi\, r^3\,e(r,\phi)},
\\ 
  &&\ve_n\,e^{i n\Phi_n}
  = -\frac{\int r\,dr\,d\phi\, r^n e^{i n\phi}\,e(r,\phi)}{\int r\,dr\,d\phi\, r^n\,e(r,\phi)},
  \quad (n>1)
\nonumber
\end{eqnarray}
where $e(r,\phi)$ is the initial energy density distribution in the plane transverse to the beam direction at the collision point $z{=}0$. Heavy-ion collision experiments at the Relativistic Heavy Ion Collider (RHIC) and the Large Hadron Collider (LHC) have established an extensive data base on corresponding anisotropies in the final momentum distributions of the emitted charged hadrons, characterized by anisotropic flow coefficients $v_n$ and their associated flow (or ``event plane") angles $\Psi_n$ \cite{Qin:2010pf,Voloshin:2008dg,Qiu:2011iv}:
\begin{equation} 
\label{eq:2}
  v_n\,e^{i n \Psi_n} =\frac
  {\int p_T\,dp_T\,d\phi_p\, e^{i n\phi_p}\,\frac{dN_\mathrm{ch}}{d\eta\,p_Tdp_T\,d\phi_p}}
  {\int p_Tdp_T\,d\phi_p\,\frac{dN_\mathrm{ch}}{d\eta\,p_Tdp_T\,d\phi_p}}.
\end{equation}
Extensive and systematic studies over the last few years (for recent reviews see 
\cite{Schenke:2011qd}) have shown that all of the measured flow anisotropies can be 
understood qualitatively, and to a large extent even quantitatively, as hydrodynamic
response to the above-mentioned initial-state fluctuations, and that the relationship between
the harmonic flow coefficients $v_n$ and corresponding eccentricities $\ve_n$ can be used
to determine \cite{Song:2010mg,Adare:2011tg,Schenke:2011bn,Qiu:2011hf} the specific shear viscosity $(\eta/s)_\mathrm{QGP}$ of the quark-gluon plasma (QGP) that (at RHIC and LHC energies) makes up the fireball medium through roughly the first half of its expansion history
\cite{Song:2007ux}. The specific shear viscosity $(\eta/s)_\mathrm{QGP}$ turns out to be
surprisingly small, $(\eta/s)_\mathrm{QGP}={\cal O}(1/4\pi)$ \cite{Song:2010mg,Adare:2011tg,Schenke:2011bn,Qiu:2011hf}, making the QGP an almost perfect fluid.

A complete theoretical analysis of all measured harmonic flow coefficients $v_1,\dots,v_6$
\cite{Adare:2011tg,ALICE:2011ab,CMS,ATLAS:2012at} is not yet available since technical shortcuts that for $v_2$ and $v_3$ allow one to sidestep the need for event-by-event hydrodynamical evolution of large numbers of fluctuating initial conditions \cite{Qiu:2011hf} fail for the higher-order harmonics \cite{Qiu:2011iv}, making their calculation numerically costly. As pointed out in \cite{Mishra:2007tw,Mocsy:2011xx}, a complete understanding of the entire spectrum of harmonic flow coefficients $v_n$ is expected to yield strong constraints on the initial conditions and dynamical evolution of heavy-ion collisions, in particular the transport coefficients of the fireball medium. The authors of \cite{Teaney:2010vd,Staig:2010pn,Nagle:2010zk,Bhalerao:2011yg,Qin:2011uw,Jia:2012ma} added that correlations between the event plane angles $\Psi_n$ of different harmonic order can yield valuable additional insights into the initial conditions. Such correlations were recently measured with good precision by the ATLAS Collaboration in Pb+Pb collisions at the LHC \cite{Jia:2012xx}. We here demonstrate that some of the measured final-state event plane correlations have a qualitatively different centrality dependence from the corresponding initial-state participant plane correlations, and that this characteristic change between initial and final state is correctly reproduced by hydrodynamic evolution. This provides additional strong support for the validity of the hydrodynamic paradigm in relativistic heavy-ion collisions. Furthermore, we show that the measured event-plane correlations are not only sensitive to the initial conditions, but also to the shear viscosity of the hydrodynamic medium, thus providing an independent constraint for this key transport coefficient.

\begin{figure*}
  \includegraphics[width=0.8\linewidth, height=0.38\linewidth]{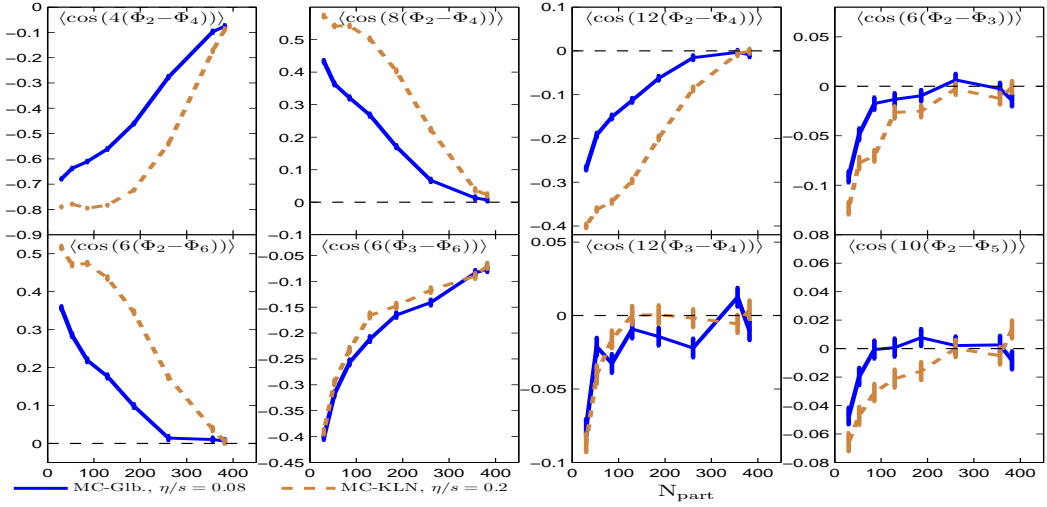}
  \caption{(Color online) Two-plane correlations $\langle\cos(jk(\Phi_n{-}\Phi_m))\rangle$, 
     where $j$ is an integer and $k$ is the least common multiple (LCM) of $n$ and $m$ 
     \cite{Bhalerao:2011yg,Jia:2012ma}, between pairs of participant-plane angles $\Phi_{n,m}$ 
     for the harmonics $(n,m)$ and multipliers $j$ studied in Ref.~\cite{Jia:2012xx}. Solid (dashed)
     lines show results for initial density profiles obtained from the MC-Glauber (MC-KLN) model.}
    \label{F1}
\end{figure*}
%
\begin{figure*}
  \includegraphics[width=0.8\linewidth, height=0.38\linewidth]{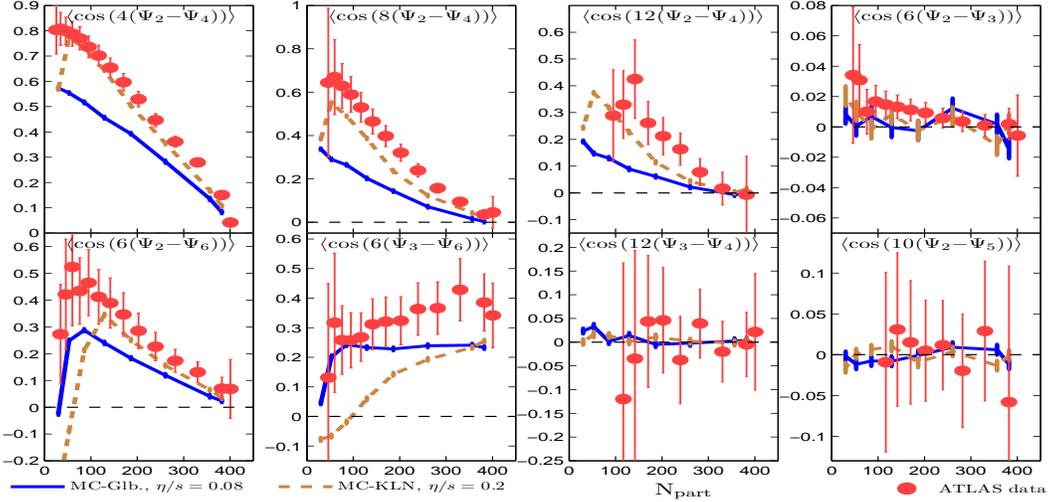}
  \caption{(Color online) Similar to Fig.~\ref{F1}, but for the corresponding final-state 
  event-plane angles 
  $\Psi_{n,m}$. Filled circles show the experimental values measured by ATLAS
  \cite{Jia:2012xx}. The MC-Glauber (solid) and MC-KLN (dashed) initial profiles used in 
  Fig.~\ref{F1} were propagated individually using viscous hydrodynamics with $\eta/s{=}0.08$
  and 0.2, respectively.}
  \label{F2}
\end{figure*}

{\sl 2.\;Methodology.}
We have used the (2+1)-dimensional code {\tt VISH2{+}1} \cite{Song:2007ux,Song:2007fn}
to evolve fluctuating initial energy density profiles for Pb+Pb collisions at $\sqrt{s}{=}2.76\,A$\,TeV event by event with viscous hydrodynamics. To explore the sensitivity to model uncertainties in the initial state, we have evolved events from two sets of initial conditions obtained from the Monte-Carlo Glauber (MC-Glb.) and the Monte-Carlo KLN (MC-KLN) models \cite{Drescher:2006pi,Hirano:2009ah}. We divided each set into centrality classes according to the number $N_\mathrm{part}$ of wounded nucleons; for each centrality class, we evolved 11,000 events from each of the two models. Model
parameters were tuned to reproduce the $p_T$ spectra and elliptic flows of unidentified charged particles and identified hadrons, as reported in \cite{Qiu:2011hf,Shen:2011eg}. This results in a specific shear viscosity $\eta/s{=}0.08$ for MC-Glauber initial conditions and the larger value $\eta/s{=}0.2$ for MC-KLN initial conditions. Both the QGP phase and the hadronic phase are evolved hydrodynamically; particle momentum distributions are calculated with the Cooper-Frye prescription, taking into account strong decays of all hadron resonances with masses up to $2.25$\,GeV. (We found, however, that the event-plane correlations discussed below are almost identical for all particle species, so including resonance decays is not essential for the present work.) From the resulting charged hadron distribution we calculate for each event the flow angles $\Psi_n$ according to
\begin{equation} 
\label{eq:3}
  v_n\,e^{i n \Psi_n} =
  \frac{\int_{0.5{<}|\eta|{<}2.5} d\eta \int_{p_\mathrm{min}} p_Tdp_T\,d\phi_p\, e^{i n\phi_p}\,\frac{dN_\mathrm{ch}}{d\eta\,p_Tdp_T\,d\phi_p}}
  {\int_{0.5{<}|\eta|{<}2.5} d\eta \int_{p_\mathrm{min}} p_Tdp_T\,d\phi_p\,\frac{dN_\mathrm{ch}}{d\eta\,p_Tdp_T\,d\phi_p}},
\end{equation}
employing the same pseudorapidity range $0.5{\,<\,}|\eta|{\,<\,}2.5$ and lower $p_T$ cutoff 
$p_T{\,>\,}p_\mathrm{min}{\,=\,}0.5$\,GeV as used in the experimental analysis \cite{Jia:2012xx}.\footnote{The ATLAS results were obtained with two independent methods: 
   (a) using a calorimetric measurement of transverse energy $E_T$ over rapidity range 
   $0.5{\,<\,}|\eta|{\,<\,}4.8$, and (b) using charged particle tracks with $p_T{\,>\,}0.5$\,GeV
   and $0.5{\,<\,}|\eta|{\,<\,}2.5$. The data from method (a) have better precision but 
   are fully compatible with those from method (b), within error bars. Since we cannot 
   simulate the calorimetric response of ATLAS theoretically, we compute the event-plane
   correlations according to method (b), but compare them in the figures to the more
   precise data obtained from method (a).}
From these event plane angles we compute for each event $\cos(k_1 \Psi_{n_1}{+}\dots{+}k_m\Psi_{n_m})$
for the two-plane ($m{=}2$) and three-plane ($m{=}3$) correlations listed in Tables 1 and 2
of Ref.~\cite{Jia:2012xx} and shown in the figures below, and then average this quantity over all events in the given centrality class. We compare these event-plane correlations with the corresponding correlations between the initial-state participant plane angles $\Phi_n$, calculated from the initial energy density profile of each propagated event according to Eq.~(\ref{eq:1}) and then averaged over events in a similar way. 

{\sl 3.\;Results.}
Figures \ref{F1} and \ref{F2} show the initial and final state two-plane correlations, for the eight
different combinations of angles and weight factors explored by the ATLAS experiment
\cite{Jia:2012xx}. Each correlation function is plotted against collision centrality, with
peripheral collisions (small $\Np$ values) on the left and central collisions (large $\Np$) on the right. Fig.~\ref{F1} shows that several of these correlations are quite sensitive to the model
used to generate the initial energy density profiles (MC-Glauber vs. MC-KLN). These model differences in the initial state manifest themselves in corresponding model differences between the final-state event-plane correlations shown in Fig.~\ref{F2}, but they are additionally modified by the different shear viscosities $\eta/s$ (0.08 and 0.2, respectively) used to evolve the initial conditions from the two models. This is most clearly seen in the ``3-6 correlation", where the two models give almost identical initial-state participant-plane correlations $\langle\cos(6(\Phi_3{-}\Phi_6))\rangle$ (second lower panel from the left in Fig.~\ref{F1}) whereas the corresponding final-state event-plane correlators $\langle\cos(6(\Psi_3{-}\Psi_6))\rangle$ exhibit significant model differences. This demonstrates the sensitivity of these event-plane correlations to the specific shear viscosity of the expanding fireball medium. 

It is worth emphasizing that several of these two-plane correlators exhibit dramatically different centrality dependences for the initial-state participant-plane and the final-state event-plane angles (see, for example, the upper left, two upper right and second lower left panels in Figs.~\ref{F1} and \ref{F2}). The difference is largest in peripheral collisions (small $\Np$). We believe that this effect is caused by a dynamical rotation of the event-plane angles during the hydrodynamic evolution \cite{unpublished}, driven by large elliptic flow in non-central collisions which leads to mode coupling between the angles $\Phi_n$ and $\Phi_{n\pm2k}$  (where $k$ is an integer and the largest coupling coefficient should correspond to $k{=}1$).\footnote{This is different from the 
   mode-coupling at freeze-out \cite{Borghini:2005kd} caused by an elliptic (quadrupole)
   deformation of the collective flow velocity appearing in the exponent of 
   the Boltzmann factor in the Cooper-Frye expression for the final particle momentum 
   distribution that couples $v_n$ with $v_{n\pm2k}$. In contrast, in the presence of 
   strong elliptic flow the non-linear hydrodynamic evolution {\it before freeze-out} leads to 
   mode-coupling between the modes $n$ and $n{\pm}2k$ ($k$ integer) for the entire 
   complex flow vector on the left hand side of Eq.~(\ref{eq:2}). We have checked 
   \cite{unpublished} that the event-plane correlations among the finally emitted particles 
   in Figs.~\ref{F2} and \ref{F4} agree qualitatively, and even almost quantitatively, with the 
   corresponding correlations between the angles associated with the harmonic coefficients 
   of the anisotropic hydrodynamic flow velocity profile along the freeze-out surface.} 
More detailed analyses will be required to confirm this conjecture. Event-plane rotation in mid-central and peripheral collisions has been suggested as the mechanism for the decorrelation of the final event-plane angle $\Psi_n$ from the initial participant-plane angle $\Phi_n$ observed in Ref.~\cite{Qiu:2011iv}, and this interpretation is consistent with the results of Ref.~\cite{Gardim:2011xv}. Being driven by mode coupling, this decorrelation depends strongly on the density and shape fluctuations in the initial state (which, after all, are the cause for non-zero odd harmonics in the final anisotropic flow); we have noticed \cite{unpublished} that the decorrelation is weaker for a viscous fluid than in ideal fluid dynamics, reflecting the dynamical smoothing by shear viscosity of initial state density fluctuations and the flows generated by them, but unambiguous evidence for a dynamical rotation of the event planes as the source of this decorrelation still needs to be established.

\begin{figure*}
  \includegraphics[width=0.8\linewidth, height=0.45\linewidth]{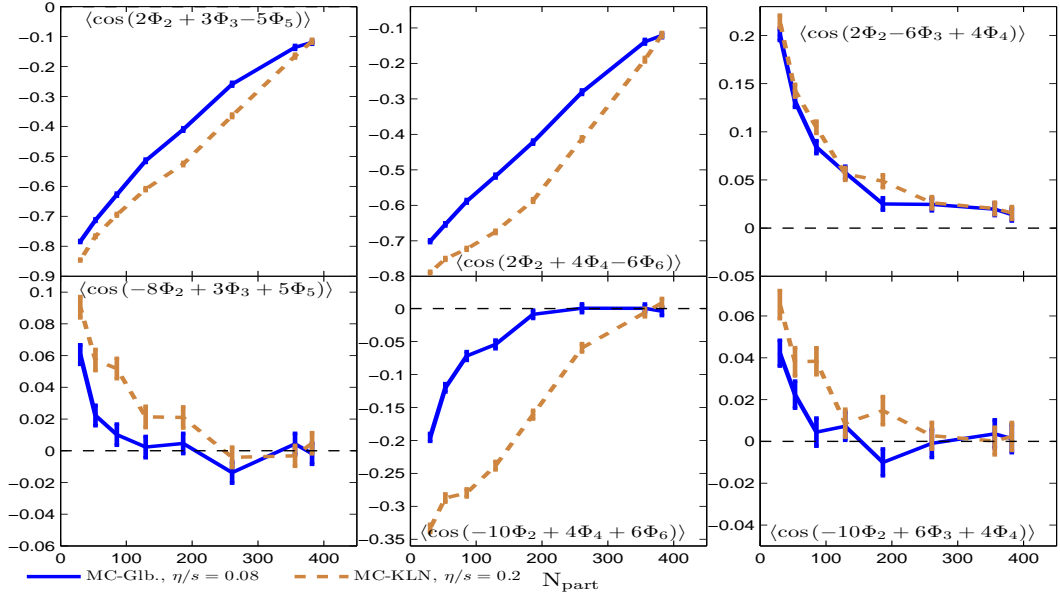}
  \caption{(Color online) Similar to Fig.~\ref{F1}, but for selected \cite{Jia:2012xx} 
  three-plane correlators 
  of the form $\langle\cos(c_l l\Phi_l{+}c_n n\Phi_n{+}c_m m\Phi_m)\rangle$, where the 
  $c_i$ are integers satisfying $c_l l{+}c_n n{+}c_m m{=}0$ \cite{Bhalerao:2011yg}.}
  \label{F3}
\end{figure*}
%
\begin{figure*}
  \includegraphics[width=0.8\linewidth, height=0.45\linewidth]{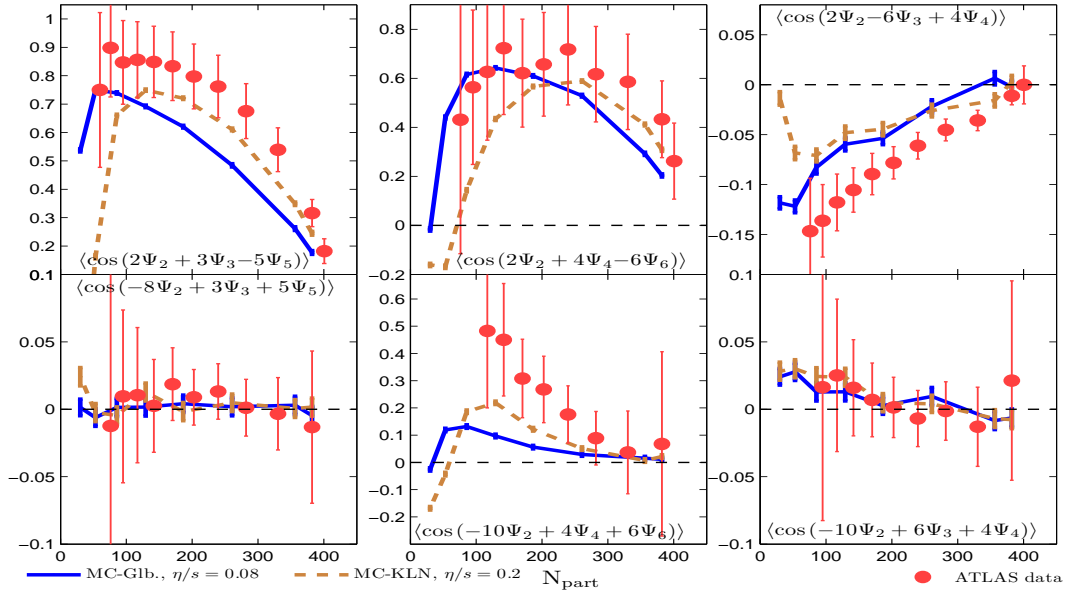}
  \caption{(Color online) The final-state event-plane correlators corresponding to the 
  initial-state correlators 
  between three participant planes of different harmonic order shown in Fig.~\ref{F3}. Solid 
  and dashed lines show results from viscous hydrodynamics with MC-Glauber and MC-KLN 
  initial conditions, evolved with $\eta/s{=}0.08$ and 0.2, respectively; filled circles show
  ATLAS data \cite{Jia:2012xx}.}
  \label{F4}
\end{figure*}

Figures~\ref{F3} and \ref{F4} show a number of three-plane correlations studied by the ATLAS experiment \cite{Jia:2012xx}, with the initial-state participant-plane correlators plotted in Fig.~\ref{F3} and the corresponding final-state event-plane correlators in Fig.~\ref{F4}, together with the experimental data. Again, we observe characteristic sign changes between several of the initial-state correlations and their corresponding final-state correlators. Even if neither of the two initial-state models (MC-Glauber and MC-KLN) reproduces the experimental data exactly, we find it impressive that the hydrodynamic model reproduces all the qualitative features of the centrality dependences of the 14 different measured event-plane correlation functions correctly: where the data show strong (weak) correlations, the same is true for the theoretical results, and where the data show correlations that increase (decrease) from peripheral to central collisions, the same holds for the theoretical predictions, without any parameter tuning. This provides very strong support for the hydrodynamic model description of the fireball evolution, from a new set of observables that is quite independent of all previously studied observables ($p_T$-spectra, anisotropic flow coefficients $v_n$, and HBT radii). 

We note that the non-linear mode-coupling first discovered in \cite{Qiu:2011iv}, and the event-plane rotations driven by this non-linear effect, are key to the qualitative agreement between theory and data in Figs.~\ref{F2} and \ref{F4}. We doubt that a similar agreement can be obtained with dynamical models that do not rely on a large degree of local thermalization in the expanding fireball, or from an approach based on linear \cite{Teaney:2010vd,Staig:2010pn,Staig:2011wj} hydrodynamic response to the initial-state density fluctuations. Inclusion of first-order non-linear terms in the hydrodynamic response \cite{Teaney:2012ke} appears to yield event-plane correlations with qualitatively similar features as shown here \cite{Teaney}, but quantitative success likely requires a numerical approach that fully accounts for the intrinsic nonlinearity of viscous hydrodynamics.

A closer look at Figs.~\ref{F1} and \ref{F3} shows that the MC-KLN model tends to produce stronger correlations between the initial-state participant-plane angles $\Phi_n$ than the MC-Glauber model. We observe that hydrodynamic evolution translates the stronger initial-state participant-angle correlations into stronger final-state event-plane correlations, even though the signs of some of the correlators featuring the strongest correlation strengths flip between initial and final state. This is especially true for the two-plane correlations shown in Fig.~\ref{F1}, while the three-plane correlators exhibit some exceptions to this ``rule" in the most peripheral collisions. The experimental data appear to prefer the stronger angle correlations in the initial profiles from the MC-KLN model, even though this model gives an elliptic-to-triangular flow ratio $v_2/v_3$ that is much larger than measured \cite{Qiu:2011hf}, caused by a larger $\ve_2/\ve_3$ ratio than in the MC-Glauber model \cite{Qiu:2011iv}. These observations show that a combined analysis of both the anisotropic flow coefficients $v_n$ and their associated flow angles $\Psi_n$ (and the correlations among them) promises to yield powerful constraints on initial state models for the fireball energy density profiles created in heavy-ion collisions. 

Even though more detailed studies will be necessary to fully explore the event-plane correlations discussed in this paper, the calculations presented here suggest that very likely neither the MC-Glauber nor the MC-KLN initial conditions will ultimately provide a quantitatively satisfactory description of the experimental data from the ATLAS Collaboration \cite{Jia:2012xx}.
How to turn the multitude of already measured and in the future measurable anisotropic flow observables (magnitudes and angles) into a focused search for the correct initial-state model is an interesting and welcome new challenge for the theory community.

{\sl Acknowledgments:}
We thank Guang-You Qin, Chun Shen, Derek Teaney and Li Yan for helpful discussions, and Jiangyong Jia and the ATLAS Collaboration for providing us with the experimental data. This work was supported by the U.S.\ Department of Energy under Grants No.~\rm{DE-SC0004286} and (within the framework of the JET Collaboration) \rm{DE-SC0004104}. We gratefully acknowledge extensive computing resources provided to
us by the Ohio Supercomputer Center.



\end{document}